\documentclass[12pt]{packages/iopart}
\usepackage{packages/iopams}
\usepackage{graphicx}
\usepackage{bm}
\expandafter\let\csname equation*\endcsname\relax
\expandafter\let\csname endequation*\endcsname\relax
\usepackage[toc,page]{appendix}
\usepackage{epstopdf}
\usepackage{cleveref}
\usepackage{siunitx}
\usepackage{float}
\usepackage[a4paper, left=4cm, right=4cm, top=2cm]{geometry}

\newpage

\begin{document}
\title{A vertical inertial sensor with interferometric readout}
\author{S.L. Kranzhoff$^{1,2,3}$, J. Lehmann$^{1,2}$, R. Kirchhoff$^{1,2}$, M. Carlassara$^{1,2}$, S.J. Cooper$^{4}$, P. Koch$^{1,2}$, S. Leavey$^{1,2}$, H. L\"uck$^{1,2}$, C.M. Mow-Lowry$^{5}$, J. W\"ohler$^{1,2}$, J. von Wrangel$^{1,2}$ and D.S. Wu$^{1,2}$}
\address{$^1$ Max Planck Institute for Gravitational Physics (Albert Einstein Institute), D-30167 Hanover, Germany}
\address{$^2$ Leibniz Universit\"at Hannover, D-30167 Hanover, Germany}
\address{$^3$ Universiteit Maastricht, 6200 MD Maastricht, Netherlands}
\address{$^4$ University of Birmingham, B15 2TT Birmingham, UK}
\address{$^5$ Vrije Universiteit Amsterdam, 1081 HV Amsterdam, Netherlands}
\ead{luise.kranzhoff@ligo.org}

\begin{abstract}
High precision interferometers such as gravitational-wave detectors require complex seismic isolation systems in order to decouple the experiment from unwanted ground motion. Improved inertial sensors for active isolation potentially enhance the sensitivity of existing and future gravitational-wave detectors, especially below 30\,Hz, and thereby increase the range of detectable astrophysical signals.
This paper presents a vertical inertial sensor which senses the relative motion between an inertial test mass suspended by a blade spring and a seismically isolated platform. An interferometric readout was used which introduces low sensing noise, and preserves a large dynamic range due to fringe-counting. The expected sensitivity is comparable to other state-of-the-art interferometric inertial sensors and reaches values of $10^{-10}\,\rm{m}/\sqrt{\rm{Hz}}$ at 100\,mHz and $10^{-12}\,\rm{m}/\sqrt{\rm{Hz}}$ at 1\,Hz. The potential sensitivity improvement compared to commercial L-4C geophones is shown to be about two orders of magnitude at 10\,mHz and 100\,mHz and one order of magnitude at 1\,Hz. The noise performance is expected to be limited by thermal noise of the inertial test mass suspension below 10\,Hz. Further performance limitations of the sensor, such as tilt-to-vertical coupling from a non-perfect levelling of the test mass and nonlinearities in the interferometric readout, are also quantified and discussed.
\end{abstract}
\pacs{04.80.Nn}
\vspace{0.5cm}
Keywords: active seismic isolation, inertial sensing, homodyne detection, interferometry, gravitational wave detection\\\\
\submitto{\CQG}
\maketitle

\section{Introduction}
\label{sec:introduction}
At measurement frequencies below 30 Hz, operating ground-based gravitational-wave detectors are limited
by control and seismic noise \cite{Martynov2016} which, with other technical noise sources like scattered light, prevents them from reaching their design sensitivities set by thermal or quantum noise. New technologies to reduce this noise are crucial for upgrades of operating facilities \cite{WhitePaper2020} as well as design studies of future detectors such as Cosmic Explorer \cite{CEHS2021} and Einstein Telescope \cite{Maggiore2020}.\\
In order to decouple the detectors from ground motion, optics in gravitational-wave detectors are suspended by multi-stage pendulums. Additionally, sensors and actuators are used in feedback control to keep the detector at its precise operating point \cite{Robertson2002,ReviewMatichard2015}. Passive and active pre-isolation of the optical tables, onto which the suspensions are placed, is used to reduce the required actuation forces and scattering, and to enable lock acquisition. For example, the internal seismic isolation (ISI) systems of Advanced LIGO (aLIGO) rely on various relative and inertial sensors to provide a seismically quiet environment \cite{Matichard2015part1,Matichard2015part2}.\\
The current seismic isolation of aLIGO is insufficiently effective at the secondary micro-seismic peak (between \SI{0.15}{\hertz} and \SI{0.35}{\hertz}) due to noise of the inertial sensors. Below \SI{0.1}{\hertz}, the performance is limited by tilt-to-horizontal coupling \cite{Matichard2016}. Although these frequencies lie outside the sensitive band of current ground-based gravitational-wave detectors, the large low-frequency motion of the suspended mirrors still couples into the measurement band through nonlinearities and potentially provides enough RMS motion to unlock the cavities \cite{Buikema2020}.\\
Several groups have developed novel inertial sensors that can be used to measure and actively stabilise tilt in order to reduce the impact of tilt-to-horizontal coupling. Some of them provide direct tilt readout, like the beam rotation sensor employed out of vacuum at aLIGO for sensor correction, which reaches a sensitivity of 1\,nrad$\sqrt{\text{Hz}}$ above 30\,mHz using a beam balance and a pair of flexures with resonance frequency 10.8\,mHz \cite{Venkateswara2014}. The rotational accelerometer ALFRA with comparable sensitivity and a total mass of 5\,kg can be mounted in an arbitrary direction and uses a walk-off sensor for the readout of a test mass (56.6\,mHz) with a mechanical quality factor ($Q$) of 8 \cite{McCann2021}. Laser gyroscopes rely on the Sagnac effect instead of using an inertial test mass and achieve an absolute rotation sensitivity in the hundred nrad/$\sqrt{\text{Hz}}$ range \cite{Korth2016,Martynov2019}. The (compact) 6D seismometers under development use a large-moment reference mass suspended from a single fused silica fibre to measure and decouple all degrees of freedom of a 3.8\,kg (3.25\,kg) test mass at the same time \cite{Mow-Lowry2019} (\cite{Ubhi2021}).
For the tilt readout, they have a predicted resonance frequency of 5\,mHz (50-100\,mHz) with a quality factor of the order $10^5$ and a sensitivity of $30 \left(400\right)\,\rm{prad}/\sqrt{\rm{Hz}}$ at 10\,mHz and $1 \left(4\right)\,\rm{prad}/\sqrt{\rm{Hz}}$ at 0.1\,Hz.\\
Additionally, several horizontal and vertical 1D inertial sensors with improved sensor resolution due to lower mechanical resonance frequencies have been developed. For example, a vertical inertial sensor based on Geometric Anti-Spring (GAS) mechanics (1\,Hz) and a capacitance displacement sensor was shown to achieve a resolution of $10^{-9}\,\rm{m}/\sqrt{\rm{Hz}}$ at 0.1\,Hz \cite{Bertolini2004} while the vertical non-magnetic optical inertial sensor (NOSE) (6\,Hz, $Q=60$) uses a leaf-spring suspended mass guided by flexures, comparable to the STS-1V \cite{Wielandt1982}, and achieves an estimated resolution of $10^{-10}\,\rm{m}/\sqrt{\rm{Hz}}$ at 1\,Hz \cite{Collette2015}. The larger Vertical Interferometric Inertial Sensor (VINS) (0.26\,Hz, $Q=30$) also uses a leaf spring suspension and has a theoretical resolution of $10^{-12}\,\rm{m}/\sqrt{\rm{Hz}}$ at 1\,Hz and $3\times10^{-10}\,\rm{m}/\sqrt{\rm{Hz}}$ at 0.1\,Hz \cite{Ding2018,DingPhD2021}. Horizontal inertial sensors like the Nikhef accelerometer with a Watt's linkage (0.45\,Hz, $Q=40$) \cite{vHeijningen2018} or the Horizontal Interferometric Inertial Sensor (HINS) with a Lehman pendulum (0.11\,Hz, $Q=15$) \cite{Ding2022} reach sensitivity levels of $10^{-12}\,\rm{m}/\sqrt{\rm{Hz}}$ at 1\,Hz and $10^{-10}\,\rm{m}/\sqrt{\rm{Hz}}$ at 0.1\,Hz. Recently, the design of a cryogenic, superconducting inertial sensor (CSIS) was published, which also uses a monolithic Watt's linkage ($Q\approx10^3-10^4$) and aims for a sensitivity of a few fm/$\sqrt{\rm{Hz}}$ at 1\,Hz \cite{VanHeijningen2022}.\\
Spatially separated pairs of 1D sensors can be used for active tilt stabilisation. Additionally, inter-table optical levers lock pairs of tables to each other for tilt stabilisation at low frequencies and allow for global tiltmeters formed by the 1D sensors placed on multiple platforms.\\
Apart from the direct tiltmeters and the GAS accelerometer, the above mentioned sensors have in common that they use an interferometric readout which generally introduces lower sensing noise than capacitive and inductive readout methods \cite{Collette2012}. Replacing the built-in readout of commercial devices by custom-made interferometers was shown to increase the sensitivity of commercial seismometers \cite{Zumberge2010} and geophones \cite{Cooper2021}.\\ 
The vertical inertial sensor presented in this paper uses a recently developed optical readout in the configuration of a phasemeter \cite{Cooper2018} to sense the relative motion between a custom-built test mass suspended by a blade spring and a pre-isolated platform. Per design, the sensor is made of vacuum-compatible components and provides adjustable passive damping of the inertial test mass. Compared to most of the sensors discussed above, the suspension has a high mechanical quality factor Q in the order of $10^{3}$ and its simple design avoids complexity associated to negative stiffness mechanisms and guiding flexures, at the price of introducing some form of cross-coupling. The sensor is expected to be limited by thermal noise of the test mass below \SI{10}{\hertz} with values of $10^{-10}\,\text{m}/\sqrt{\text{Hz}}$ at 0.1\,Hz and $10^{-12}\,\text{m}/\sqrt{\text{Hz}}$ at 1\,Hz.\\
In this paper, we present the mechanical design of the sensor in \cref{sec:sensor-design} and its expected sensitivity in \cref{subsubseb:noise-calculation}. We use the seismically quiet environment of the passively and actively isolated platform of the AEI 10\,m Prototype facility \cite{Kirchhoff2020} and its various motion and temperature sensors to measure the noise performance in \cref{subsubsec:noise-measurement} and quantify effects of temperature drifts on the dynamic behaviour of the sensor in \cref{subsec:ttv-coupling} and its calibration in \cref{subsec:nonlinearities}.

\section{Sensor Design}
\label{sec:sensor-design}
\begin{figure}[h]
\centering
\includegraphics[trim = 0cm 0cm 0cm 0cm, clip, width=15cm]{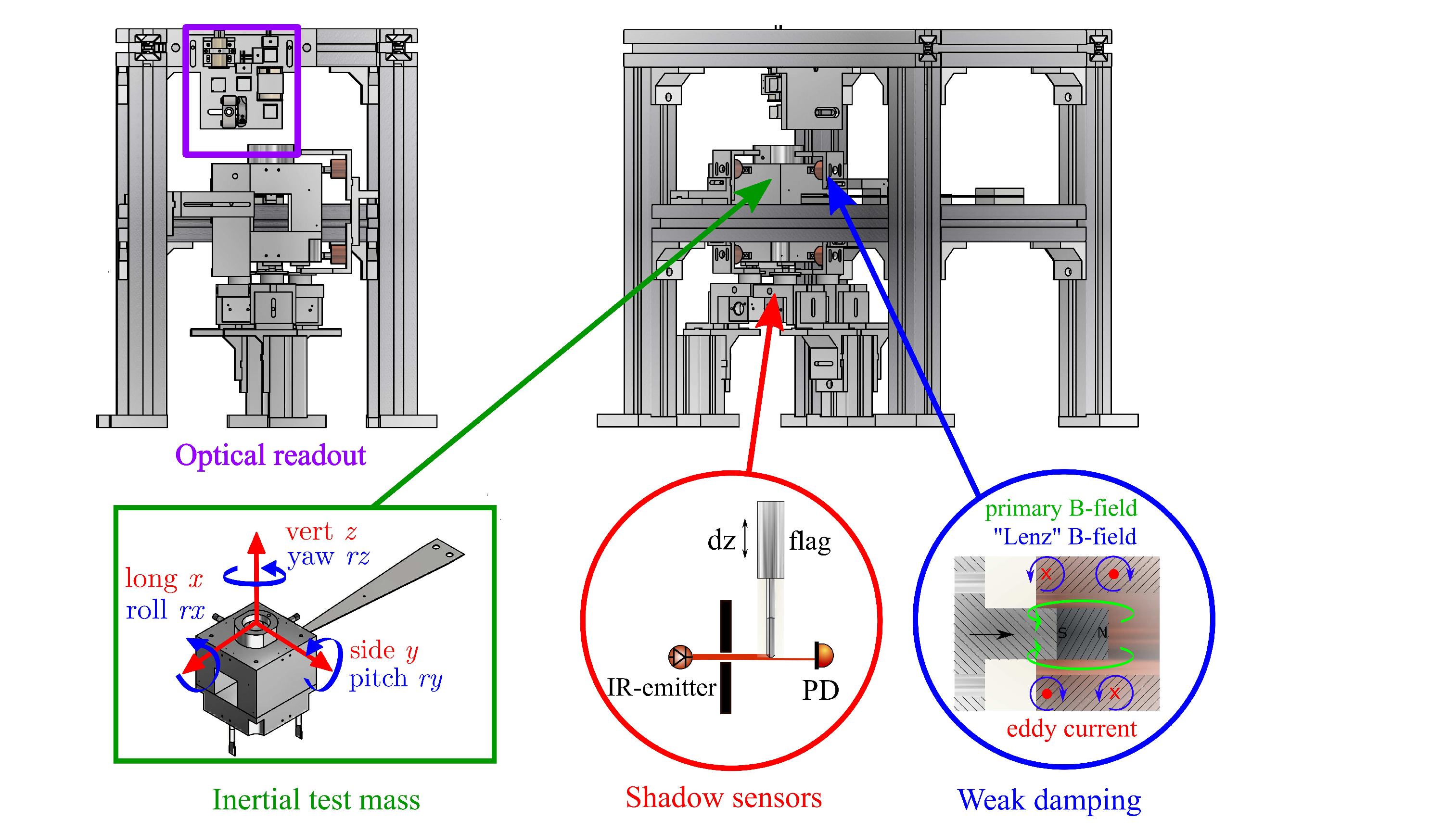}
\caption{Schematic of the mechanical sensor design. The interferometer (purple) reads out differential motion between the inertially suspended test mass (green) and the suspension cage, which is stiffly connected to the table. Shadow sensors (red) serve as second readout for test mass motion to measure and verify the suspension properties. Eddy current dampers (blue) attenuate test mass motion to reduce nonlinear effects.}
\label{fig:sensor-design_overview}
\end{figure}

An overview of the different components of the prototype sensor design is shown in \cref{fig:sensor-design_overview}.
The test mass suspension is realised by a maraging steel C250 spring blade which is pre-bent to be flat under a load $m$ of \SI{1.2}{\kilo\gram}. The test mass is levelled and aligned in rotation by placing additional mass on top of the cube. By design, the blade spring and mass assembly has a vertical resonance frequency $f_{0,\rm{z}}$ of \SI{1}{\hertz}.\\
The readout instrument is an adapted version of the Homodyne Quadrature Interferometer (HoQI) presented in \cite{Cooper2018}. It uses a pair of orthogonal polarisation states of a laser beam to monitor differential arm length changes over multiple optical fringes. The corresponding optical phase is calculated via the arctangent of the two orthogonal quadrature signals at the interferometer output which allows a linearisation of the signal compared to standard Michelson interferometers.
Geometric shadow sensors \cite{AstonPhD2011} are used as additional readout to measure and verify the suspension properties. The mechanical quality factor of the suspension is measured with shadow sensors to be $Q_{\text{s}}>2500$ in air. The next resonance frequencies of the suspension were found to be the roll resonance $f_{0,\rm{rx}}$ at \SI{4.55}{\hertz} and the pitch resonance $f_{0,\rm{ry}}$ at \SI{10.76}{\hertz}. Both are close in frequency to the vertical resonance and lie in the target frequency band of active seismic isolation, potentially disturbing the vertical readout and complicating the design of feedback loops. However, measurements of actuated transfer functions with an active platform show that the coupling of platform motion in roll and pitch into the interferometric readout is at least two orders of magnitude lower than the coupling of vertical motion.\\
For a first test of the sensor, eddy current dampers are used as passive velocity-dependent damping method for the inertial test mass. In total, four eddy current dampers are installed leading to a measured viscous damping coefficient $b$ of \SI{0.19}{\kilo\gram\per\second}. The damping strength can be chosen via the dimensions of the copper tubes as well as dimensions and material of the magnets moving within the tubes. Longitudinal and lateral motion of the magnet inside the copper tube are damped equally leading to an increased thermal noise of the test mass corresponding to a lowering of the mechanical quality factor to $Q_{\rm{v}} \approx 44$. In this prototype version of the sensor, the increase in thermal noise was accepted in order to reduce the effect of nonlinearities in the interferometric readout by reducing the RMS motion of the inertial test mass. Due to the different frequency dependencies of structural and viscous thermal noise, structural thermal noise of the blade material still limits the sensor performance below a certain frequency set by the damping strength.
\begin{figure}[h]
\centering
\includegraphics[trim = 0cm 0cm 0cm 0cm, clip, width=9cm]{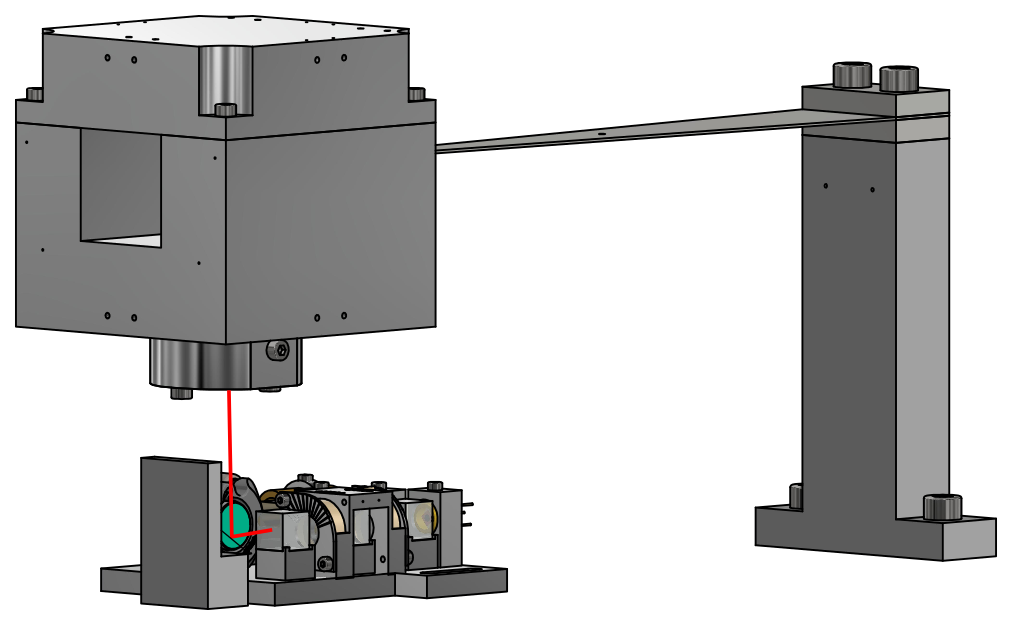}
\caption{A possible redesign for the sensor with a total mass of \SI{1.75}{\kilo\gram}.}
\label{fig:sensor_light-design}
\end{figure}\\
For the prototype version of the sensor, a cage of dimensions \SI{320}\times\SI{230}\times \SI{310}{\milli\metre} houses all sensor components and the sensor has a total mass of \SI{6}{\kilo\gram}. This mass can be reduced to a mass comparable with Sercel L-4C geophones (about \SI{2}{\kilo\gram}) when removing shadow sensors and damping, and optimizing the design of the cage. A possible light-weight design with a total mass of \SI{1.75}{\kilo\gram} is shown in \cref{fig:sensor_light-design}.

\section{Sensor Performance}
\label{sec:sensor-performance}
This section describes the effects that limit the performance of the novel vertical inertial sensor in terms of noise (\cref{subsec:sensitivity}), mechanical tilt-to-vertical coupling (\cref{subsec:ttv-coupling}) and nonlinearities in the interferometric readout (\cref{subsec:nonlinearities}).

\subsection{Noise Sources}
\label{subsec:sensitivity}
\begin{figure}[h]
\centering
\includegraphics[trim = 0cm 0cm 0cm 0cm, clip, width=13cm]{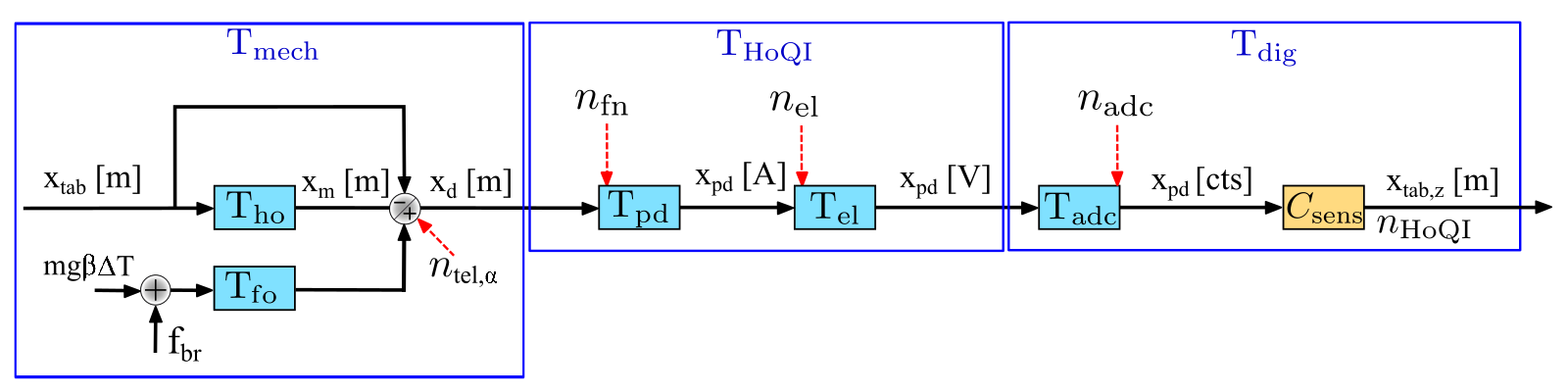}
\caption{Block diagram of the HoQI-based inertial sensor including transfer functions of individual components $\mathrm{T}_{i}$, signals $x_{i}$ and noise contributions $n_{i}$. The photodiode response $\mathrm{T}_{\rm{pd}}$, transfer functions of the readout electronics $\mathrm{T}_{\rm{el}}$ and the analogue-to-digital converter $\mathrm{T}_{\rm{adc}}$ are frequency-independent in the frequency band of interest. The block $C_{\rm{sens}}$ represents the mathematical functions applied digitally to the three photodiode signals to calculate the optical phase of the interferometer.}
\label{fig:noise-paths}
\end{figure}

The block diagram in \cref{fig:noise-paths} indicates the relevant transfer functions $\mathrm{T}_i$ for sensor calibration and noise contributions $n_i$ for a sensitivity estimation. The sensor reads out the relative (vertical) motion $x_{\rm{d}}$ between the table $x_{\rm{tab}}$ and the inertial test mass $x_{\rm{m}}$. The inertial test mass has the mechanical transfer function of a harmonic oscillator $\mathrm{T}_{\text{ho}}$ with a corresponding force response transfer function $\mathrm{T}_{\text{fo}}$. All noise contributions can be projected into equivalent table displacement via plant inversion, i.e. multiplying by the inverse of
\begin{equation}
\mathrm{T}_{\text{ho}}-1 
= \frac{x_{\text{d}}}{x_{\text{tab}}} 
= \frac{\omega^2}{\omega_{0}^2\left(1+\mathrm{i}\frac{1}{Q_{\text{s}}}\right)-\omega^2+\mathrm{i}\frac{b}{m}\omega}
=m\omega^2\mathrm{T}_{\text{fo}}\,.
\end{equation}
In the following, the calculation of all relevant noise sources is summarised leading to the expected displacement sensitivity of the inertial sensor shown in \cref{fig:noise-budget}.

\subsubsection{Noise Calculation}
\label{subsubseb:noise-calculation}
\begin{figure}[h]
\centering
\includegraphics[trim = 0cm 0cm 0cm 0cm, clip, width=13cm]{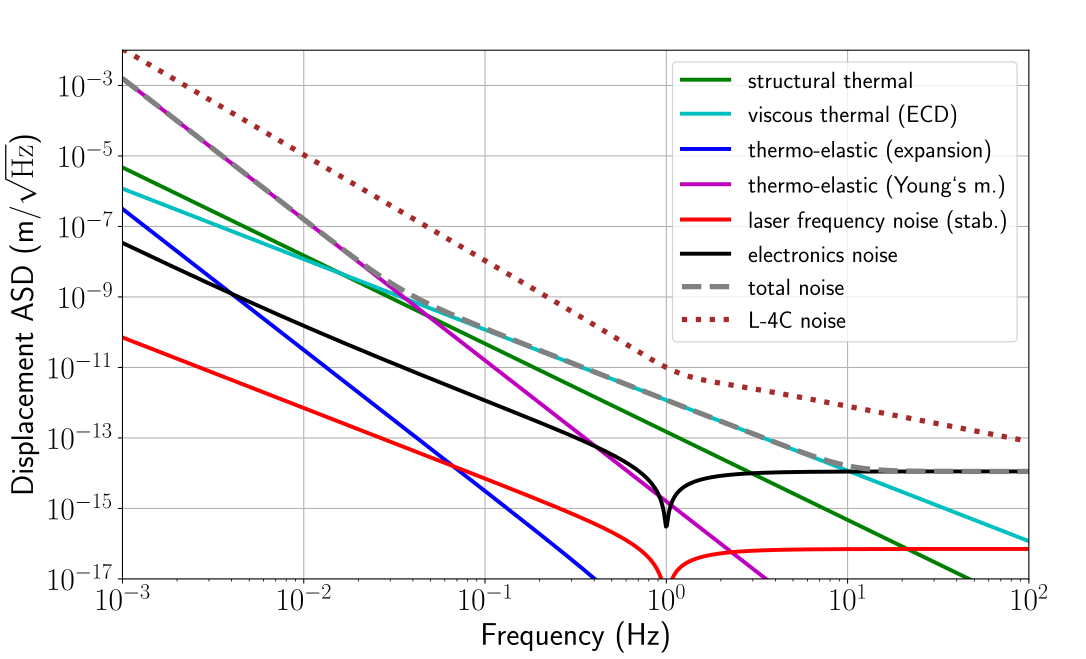}
\caption{Modeled noise budget for the novel vertical inertial sensor depicted as displacement amplitude spectral density (ASD). The total sensor noise (dashed grey) is modeled to be dominated by thermoelastic noise due to temperature-induced change in Young's modulus below \SI{30}{\milli\hertz} (solid purple), by viscous thermal noise due to eddy current damping (ECD) between \SI{30}{\milli\hertz} and \SI{10}{\hertz} (solid cyan).}
\label{fig:noise-budget}
\end{figure}

Different thermal noise sources, all based on the fluctuation-dissipation theorem \cite{FDT1952}, need to be considered. Brownian noise manifests in a fluctuating force \cite{Saulson1990}, 
\begin{equation}
    f_{\text{br}}=\sqrt{4k_{\text{B}}T\left(b+\frac{m\omega_{0}^2}{Q_{\text{s}}\omega}\right)}\,,
\end{equation}
and combines structural thermal noise due to internal losses in the maraging-steel blade material and viscous thermal noise due to external velocity damping.\\
Additionally, two sources of thermoelastic noise need to be considered. One originates from thermal expansion of the aluminium and stainless steel components causing the HoQI arm lengths to be temperature dependent, $L_{i}\left(\Delta T \right) = L_{i0} \left(1+\alpha_{j}\Delta T \right)$, and leading to a displacement noise $n_{\text{tel},\alpha}$. It couples via the differential arm length change $\Delta L_{\rm{yz}}\left(\Delta T\right)$ with the $y$- and $z$-arm lengths given by,
\begin{eqnarray}
\label{eq:arm-length}
L_{\rm{y}}\left(\Delta T\right)&=L_{\rm{y0}}\left(1+\alpha_{\rm{al}}\Delta T\right)\,,\\
L_{\rm{z}}\left(\Delta T\right)&=L_{\rm{z0}}\left(1+\alpha_{\rm{al}}\Delta T\right)+z_{\rm{bl}}\alpha_{\rm{m}}\Delta T\,,
\end{eqnarray}
where $z_{\rm{bl}}=0.8\,$mm denotes the thickness of the blade spring at room temperature. With the thermal expansion coefficients $\alpha_{\rm{al}}=2.3\times 10^{-5}\,\rm{K}^{-1} $ for aluminium \cite{Hidnert1952} and $\alpha_{\rm{m}}=1\times 10^{-5}\,\rm{K}^{-1}$ for maraging steel \cite{Gerlich1990} and a measured arm length mismatch of $\Delta L_{\rm{yz,dc}}=L_{\rm{y0}}-L_{\rm{z0}}=0.2\,$mm, this results in a temperature-based differential arm length change of
\begin{eqnarray}
\label{eq:armlength-temperature}
 \frac{d\Delta L_{\rm{yz}}}{d\Delta T}&=\frac{d}{d\Delta T}
 \left(L_{\rm{y}}\left(\Delta T\right)-L_{\rm{z}}\left(\Delta T\right)-\Delta L_{\rm{yz,dc}}\right) \nonumber \\
 &=\Delta L_{\rm{yz,dc}} \alpha_{\rm{al}} + z_{\rm{bl}}\alpha_{\rm{m}}
 \label{eq:ten-expansion} \\
 &\approx 1.26\times 10^{-8}\,\frac{\rm{m}}{\rm{K}}\,. \nonumber
\end{eqnarray}
The other thermoelastic noise source originates from the temperature-induced relative change in Young's modulus $\beta_{\text{m}}$, which amounts to $2.54\times 10^{-4}\,\rm{K}^{-1}$ for the maraging-steel spring blade \cite{Cordero2000,Ledbetter1977}. It influences the spring constant linearly via $k\left(\Delta T\right)= k\left(0\right)\left(1+\beta_{\rm{m}}\Delta T\right)$ and its effect is covered by a modification of the harmonic oscillator equation of motion reading
\begin{equation}
0 = - mg + k\left(\Delta T\right)\left(z_{\rm{wp}}-z\right)\,,
\end{equation}
which results in a force noise $mg\beta_{\text{m}}\Delta T$. Assuming $z_{\rm{wp}}=mg/k\left(0\right)$ for the working point leads to the temperature-based change of equilibrium position of
\begin{eqnarray}
\label{eq:nonlinear-tel}
\frac{dz}{d \Delta T}
=\frac{g\beta_{\rm{m}}}{\omega_{0}^{2}}
\approx 6.31\times 10^{-5}\,\frac{\rm{m}}{\rm{K}}\,,
\end{eqnarray}
below the test mass resonance frequency and to linear order in $\beta_{\text{m}}$. The Young's modulus coupling factor in eq. \ref{eq:nonlinear-tel} was confirmed with measurements using the spectrum of custom-made in-vacuum temperature sensors whose signals were coherent with the HoQI signal in the few mHz-regime. Similar considerations were made in \cite{OteroPhD2009,BergmannPhD2018}.\\
The good agreement of theory and measurement could be partly caused by the simple suspension geometry. Replacing the spring material with an Elinvar alloy to compensate this effect for future versions of the sensor seems promising. In order to increase the time constant of the temperature fluctuations, the sensor could be enclosed in a gold-plated metallic heat shield.\\ 
For an estimation of the thermoelastic noise contribution in the inertial sensor noise budget, a spectrum is taken from data of the temperature sensors in vacuum, see \ref{ap:temp}. The temperature measurement is sensor-noise dominated for frequencies above a few mHz. The meaningful data were extrapolated by applying a fit curve with $1/f^2$ slope to the measured value at 1\,mHz.\\
Regarding laser noise, mainly laser frequency noise $n_{\rm{fn}}$ couples into the readout. Common-mode laser intensity noise can be sufficiently subtracted in the digital processing assuming that offsets due to gain differences between individual photodiodes vary by less than a few percent. For the noise budget in \cref{fig:noise-budget}, the frequency noise of an Innolight Prometheus laser is assumed to be $\delta f = 100\,\rm{Hz}/\sqrt{\rm{Hz}}$ above 10\,mHz \cite{Scholz2009}.\\
 For the calculation of thermoelastic noise as well as laser frequency noise, the residual arm length mismatch $\Delta L_{\rm{yz,dc}}$ is assumed to be \SI{0.2}{\milli\metre} which was experimentally achieved. \cref{fig:noise-budget} shows that, even for a large arm length mismatch, laser frequency noise does not limit the sensor sensitivity.\\
The readout electronics were based on a design provided by the University of Birmingham and components were chosen not to limit the sensor performance in the frequency band of interest. As apparent in \cref{fig:noise-budget}, electronics noise, dominated by Johnson noise of the resistors as well as voltage and current noises of the operational amplifiers, are modeled to limit the sensor performance above 10\,Hz. Readout noise is equal to shot noise for an optical power of \SI{0.09}{\milli\watt} hitting individual photodiodes. Since the readout would saturate at a power of \SI{0.49}{\milli\watt}, both noise contributions are of comparable magnitude. For simplicity, \cref{fig:noise-budget} displays only electronics noise. The electronics noise can be modelled or measured by injecting offsets that resemble a certain operating point of the interferometer. With that, the resulting equivalent displacement noise of the electronics is found to limit the sensor at high frequencies and to be in the order of $10^{-14}\,\rm{m}/\sqrt{\rm{Hz}}$. Although electronics noise rises towards lower frequencies below the test mass resonance, it is not limiting the sensor performance for any frequency below \SI{10}{\hertz}.
\\
In \cref{fig:noise-budget}, the expected noise performance of the inertial sensor is compared to the noise performance of L-4C geophones \cite{KirchhoffHuddle2017} which are commonly used as horizontal and vertical inertial sensors in the AEI 10\,m Prototype and other facilities. The total noise of the inertial sensor presented in this paper is calculated by the squared sum of the individual uncorrelated noise terms projected into equivalent table displacement.\\
The result shows that for frequencies below \SI{30}{\milli\hertz} the sensor is expected to be limited by changes in Young's modulus due to temperatures fluctuations. At frequencies from \SI{30}{\milli\hertz} to about \SI{10}{\hertz}, the sensor is calculated to be limited by viscous thermal noise due to the eddy current damping mechanism. Here it is assumed that at these frequencies, the analogue-to-digital converter noise is suppressed below electronics noise through the use of whitening filters. Assuming the depicted noise performance, the novel vertical inertial sensor is expected to be two orders of magnitude more sensitive than currently used L-4C geophones at \SI{10}{\milli\hertz} and at \SI{100}{\milli\hertz}. At the resonance of \SI{1}{\hertz}, the performance is improved by one order of magnitude.

\subsubsection{Noise Measurement}
\label{subsubsec:noise-measurement}
Huddle tests are a method to measure the sensor noise and to verify the calculated sensitivity in frequency regions where large foreground signals mask the noise. For example, for L-4C and L-22D geophones, results of huddle tests were shown to agree with calculated noise budgets \cite{KirchhoffHuddle2017}.\\
In general, the inertial sensor signal is composed of a contribution from detected motion  and sensor noise, which can be separated to a certain degree by employing reference sensors located close to the sensor to be characterized \cite{Holcomb1989}. The detected motion is understood to be coherent between all sensors and is subtracted so that the incoherent part of the signal, i.e. the sensor noise, remains.
The method of subtraction in the frequency domain is based on \cite{Allen1999} and the script used for the huddle test results in this paper can be found in the supplementary material of \cite{KirchhoffHuddle2017}.\\
The huddle test of the HoQI-based inertial sensor was conducted at the AEI \SI{10}{\metre} Prototype facility in Hanover, Germany \cite{Gossler2010}. For the measurements, the sensor was placed inside the vacuum system of the prototype detector on top of a passively and actively isolated platform \cite{Kirchhoff2020}. In the frequency region of interest, the sensor noise is not directly measurable because the achievable residual table motion lies above the calculated noise performance of the HoQI-based inertial sensor.\\
The huddle test was performed on an eight-hour-long time series of data. The measurement was done overnight at the weekend where seismic motion due to anthropogenic activities is expected to be low. Additionally, optical table motion was suppressed using active control. Angular motion was measured and stabilised using optical levers as in-loop sensors. As reference sensors, three vertical and three horizontal L-4C geophones, two optical levers and a broadband Streckeisen STS-2 seismometer measuring the ground motion were used. The approach taken was broadly similar to \cite{KirchhoffHuddle2017}.
\begin{figure}[h]
\centering
\includegraphics[trim = 0cm 0cm 0cm 0cm, clip, width=13cm]{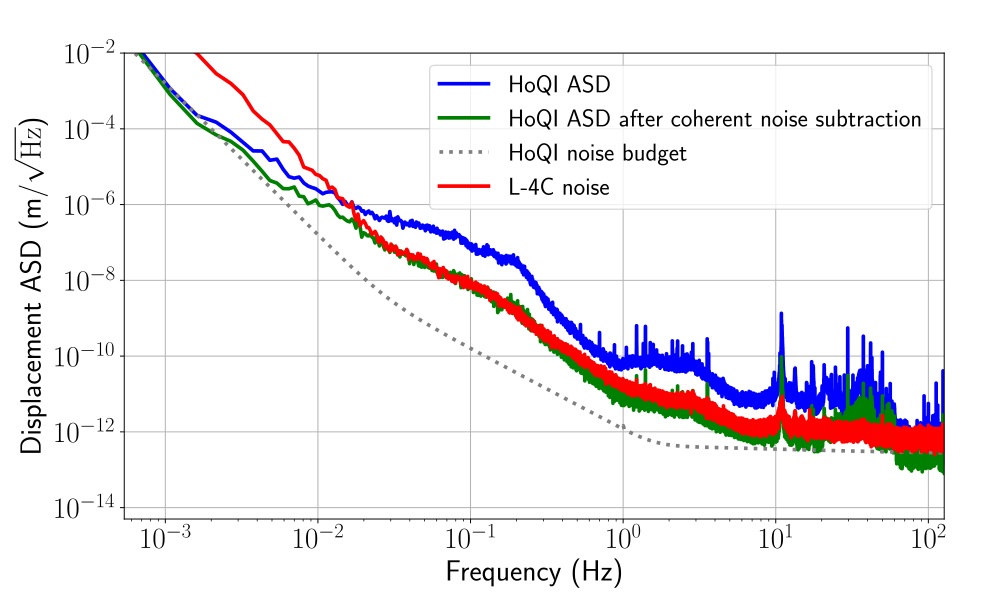}
\caption{Result of a huddle test for the novel vertical inertial sensor carried out in vacuum using a seismically isolated platform.}
\label{fig:huddle-result}
\end{figure}\\
The result is shown in \cref{fig:huddle-result}. Above \SI{20}{\milli\hertz}, the multichannel coherent subtraction is limited by the noise of the reference L-4C sensors. The noise of the HoQI-based inertial sensor cannot be proven to lie below L-4C noise in this frequency region since the measured HoQI amplitude spectral density lies above L-4C noise and the geophones are the most sensitive reference sensors in this frequency region. Below \SI{20}{\milli\hertz}, the HoQI signal already lies below L-4C noise so that geophone signals are not coherent with the HoQI signal. Differences between the HoQI amplitude spectral density and the noise curve are caused by subtraction of coherent parts with the optical levers, the STS-2 seismometer ground motion signals and tilt-to-horizontal coupling to the horizontal L-4C geophones. At frequencies below a few mHz, the result of the huddle test approximately matches the noise budget in \cref{fig:noise-budget} but has a slightly different slope.

\subsection{Tilt-to-vertical Coupling}
\label{subsec:ttv-coupling}
\begin{figure}[h]
\centering
\includegraphics[trim = 0cm 0cm 0cm 0cm, clip, width=13cm]{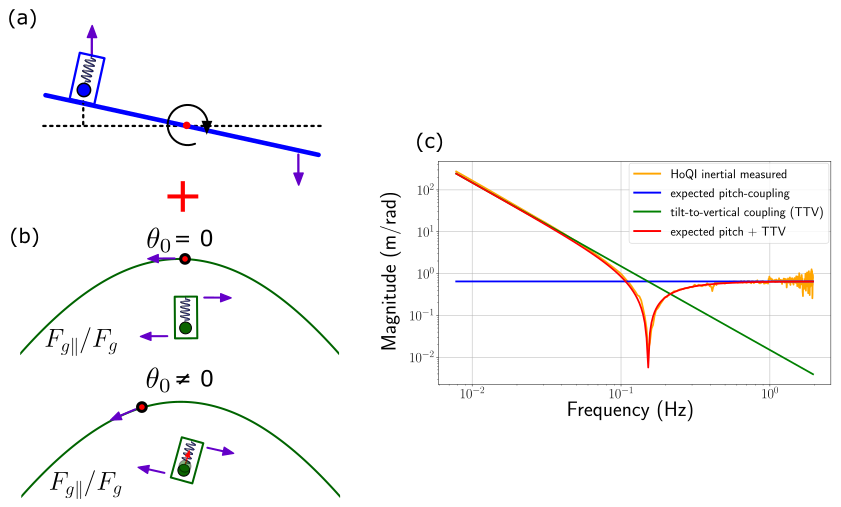}
\caption{The total coupling of table tilt into the sensor signal (red) is the sum of a (a) position-based coupling (blue) and (b) tilt-to-vertical coupling (green), the latter scaling with the sensor's angular misalignment $\theta_{0}$ with respect to gravity. (c) shows the transfer function for vertical inertial motion measured with the HoQI for pitch actuation (orange). It is reproducible from the expected position-based, frequency independent pitch coupling (blue) and tilt-to-vertical coupling (green).}
\label{fig:ttv-schematic}
\end{figure}

At frequencies below 150\,mHz, the coupling of table tilt to the inertial sensor signal is dominated by a tilt-to-vertical (TTV) coupling mechanism which is $180^{\circ}$ out of phase with respect to the tilt coupling $\mathrm{T}_{\text{ry}}^{\text{HoQI}}$ resulting from sensor positioning and suspension-based cross coupling. A similar effect has been reported in \cite{Zhao2020} for the control of a single-degree-of-freedom seismic isolation system. The coupling strength shows a frequency dependence $\sim 1/\omega^{2}$ analogously to the well-known tilt-to-horizontal coupling of horizontal inertial sensors and scales with the static misalignment $\theta_{0}$ of the spring blade with respect to gravity. The coupling mechanism is based on the fact that the signal of a statically tilted sensor $x_{\text{sensor}}$ does not only detect a vertical motion $x_{\text{vert}}$ but will also measure a fraction of horizontal table motion $x_{\text{hor}}$ including tilt-to-horizontal coupling 
\begin{equation}
x_{\rm{sensor}}=x_{\rm{vert}}\cdot\cos\left(\theta_{0}\right)
+\left(x_{\rm{hor}}+{\delta\theta}\cdot\frac{g}{\omega^2}\right)\cdot\sin\left(\theta_{0}\right)\,,
\end{equation}
where $g$ is the gravitational constant and $ \delta\theta $ the dynamic tilt of the table.
This principle is illustrated in \cref{fig:ttv-schematic}.
\\
TTV coupling spoils the sensor performance at low frequencies since it creates an unwanted signal shading the signal caused by real vertical motion. With the position-based signal from tilt, $\delta\theta\cdot\mathrm{T}_{\rm{ry}}^{\rm{HoQI}}\cdot\cos\theta_0$, the crossover frequency $\omega_{\rm{dip}}$ of both coupling mechanisms in the measured actuated pitch transfer function is
\begin{equation}
\label{eq:crossover-freq}
\centering
    \omega_{\rm{dip}}^2=\frac{g\cdot\sin\theta_0}{\mathrm{T}_{\rm{ry}}^{\rm{HoQI}}\cdot\cos\theta_0}\,.
\end{equation}
From this, the angle of static misalignment for the vertical inertial sensor is calculated. For the vertical inertial sensor presented in this paper, the effect on the measured actuated pitch transfer function corresponds to a static misalignment of $\theta_0\approx 2.87^{\circ}$. Lowering the crossover frequency to 10\,mHz would require a static misalignment of $\theta_0<0.012^{\circ}$. This requirement is a challenge since in the current configuration vertical motion of the test mass is fundamentally coupled to angular misalignment of the same. While the buoyancy effect causing a misalignment of $0.025^{\circ}$ can be pre-compensated before pumpdown, the coupling of temperature fluctuations with about $0.016^{\circ}$/K would motivate a change in the suspension. However, even with a configuration where the test mass moves on a straight line along the vertical axis, a careful levelling of the sensor frame is needed to achieve a sufficient suppression of coupling to table tilts. Suitable alternative suspension designs are discussed in \ref{ap:susp-designs}. If the spring material would be replaced with an Elinvar alloy to suppress thermoelastic noise, this would also reduce the temperature dependence of the sensor alignment.

\subsection{Nonlinearities in the Interferometric Readout}
\label{subsec:nonlinearities}
Optimally, the quadrature readout of the HoQI linearises the signal compared to standard Michelson interferometers. However, there are various errors caused by, for instance, imperfections of optics, misalignment of the HoQI or manufacturing differences in the photodiodes and the readout electronics. These errors are classed into offset, quadrature and gain errors \cite{Watchi2018} and lead to the calculated phase being nonlinear to the differential arm length change of the interferometer. Nonlinear upconversion of signals or noise is visible in the measured amplitude spectral density as higher order harmonics and a shelf-like feature. Consequences of this effect on the HoQI readout have been investigated in \cite{CooperPhD2019} and post-processing techniques, e.g. an ellipse fitting technique \cite{Heydemann1981}, can be used to significantly reduce them.\\
For the vertical inertial sensor, the nonlinear dependence of the inertial sensor calibration on the operating point of the HoQI has been analysed. For this investigation, a small single-frequency excitation of the active platform was performed continuously over a period of 12\,h, during which the HoQI passes through all operating points due to temperature drifts. The operating point (optical phase) of the interferometer at a certain time was determined by calculating the relative response between the vertical L-4C geophone signal and the HoQI-based inertial sensor to the excitation, provided that the coherence of the sensor signals with the excitation was high.
\begin{figure}[h]
\centering
\includegraphics[trim = 0cm 0cm 0cm 0cm, clip, width=13cm]{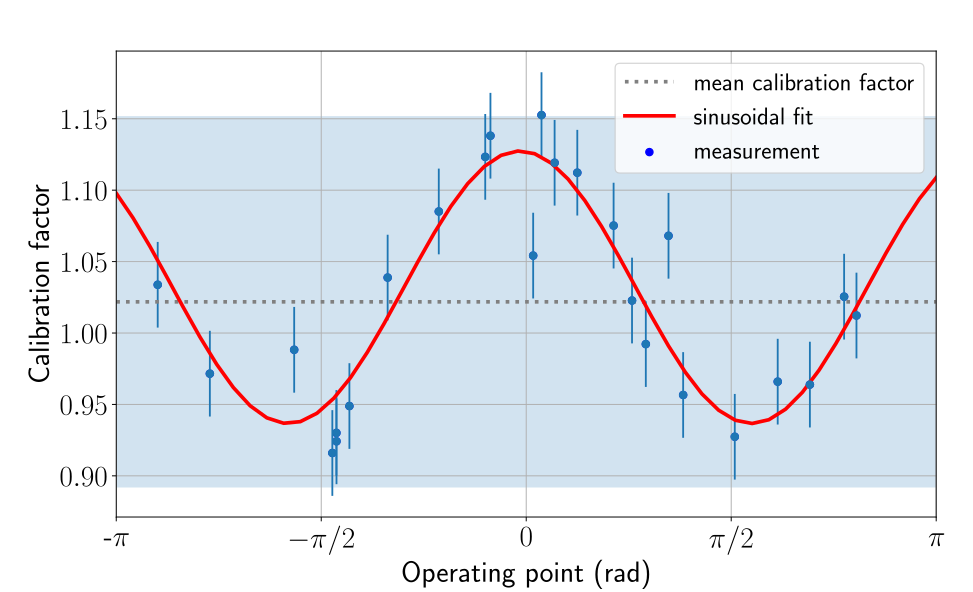}
\caption{Measured nonlinear dependence of the calibration on the operating point of the HoQI (blue
dots). The error bars are set by the deviations occurring between averages in the
measurement. The data points are fitted (solid red) to a function given in \cref{eq:nonlin-calib-fit}. In the measurement, deviations of up to 13\,\% from the mean calibration occurred (blue-shaded area).}
\label{fig:calib-drifts}
\end{figure}\\
The results of the measurement are displayed in \cref{fig:calib-drifts}. The calibration shows an oscillatory dependence on the optical phase. It is therefore sensible to use a fit of the form
\begin{equation}
C_{\rm{z}}\left(\phi_{\rm{opt}}\right) = C_{0} + C_{\rm{pk}} \cdot \sin\left(b\cdot\phi_{\rm{opt}}+\phi_{0}\right)\,.
\label{eq:nonlin-calib-fit}
\end{equation}
The numerical parameters are
\begin{itemize}
\item the mean value of the relative calibration $C_{0}=1.03$\,,
\item the peak amplitude of the oscillation $C_{\rm{pk}} = 0.095$\,,
\item the frequency of the oscillation $b = 1.77$\,,
\item the phase offset $\phi_{0} = 1.66 \approx 0.53\,\pi$\,.
\end{itemize}
The result describes a cosine oscillation ($\phi_{0} \approx \pi/2$) around the value $C_{0}=1.03$, which corresponds well to the measured mean vertical coupling factor of the inertial sensor. The oscillation amplitude $C_{\rm{pk}}=0.095$ determines the potential calibration error due to disregard of the operating point which can drift due to temperature fluctuations. \\
In order to prevent harmonics from spoiling the inertial sensing, this high nonlinearity needs to be accounted for before the damping of the test mass can be removed.

\section{Discussion}
\label{sec:discussion}
In this paper, the HoQI readout of a custom-built suspended test mass was presented with the aim of designing a vertical inertial sensor with a high displacement sensitivity comparable to commercial broadband seismometers and a low mass comparable to commonly used geophones. The calculated sensitivity promises two orders of magnitude improvement compared to L-4C geophones at 10\,mHz and 100\,mHz and one order of magnitude at 1\,Hz but a huddle test only verified a lower noise of the HoQI-based inertial sensor below 20\,mHz and an equal noise level above 20\,mHz with values of $10^{-8}\,\rm{m}/\sqrt{\rm{Hz}}$ at 100\,mHz and $10^{-11}\,\rm{m}/\sqrt{\rm{Hz}}$ at 1\,Hz. The results presented here show that the sensitivity of a HoQI-based inertial sensor is as good as an L-4C geophone, as an upper limit. However, multiple HoQI-based sensors would be required as reference sensors in order to measure the actual noise performance of these sensors. With the calculated sensitivity, the sensor would be in the same order of magnitude as for other state-of-the art inertial sensors with custom mechanics and interferometric readout. For a future version of the sensor, decreasing thermal noise by removing eddy current damping or replacing it by active damping or data processing techniques should be reevaluated depending on the intended use case. \\
The simplicity of the suspension design leads to three key challenges to overcome, namely low resonance frequencies in the vertical rotational degrees of freedom roll and pitch, the fundamental coupling between vertical and pitch motion of the test mass, and the challenge of aligning the test mass so that it moves in parallel to gravity.
The coupling strength of buoyancy effects and temperature fluctuations on the test mass equilibrium position and its dynamics are partly driven by the fundamental coupling of vertical motion to the angular alignment of the test mass and partly by the choice of the blade spring material. The required alignment accuracy to sufficiently suppress tilt-to-vertical coupling is a challenge with the current suspension design, especially with regard to temperature fluctuations which are difficult to predict.

\ack
The authors are grateful for the support from the International Max Planck Research School (IMPRS) on Gravitational Wave Astronomy, QUEST, the Center for Quantum Engineering and Space-Time Research and Quantum Frontiers and the Deutsche Forschungsgemeinschaft (DFG, German Research Foundation) under Germany’s Excellence Strategy – EXC-2123 QuantumFrontiers – 390837967.\\
The authors would like to thank D. Hoyland from University of Birmingham for providing schematics for the HoQI readout electronics.\\\\

\newpage
\appendix

\section{Modelling temperature fluctuations}
\label{ap:temp}
\begin{figure}[h]
\centering
\includegraphics[trim = 0cm 0cm 0cm 0cm, clip, width=13cm]{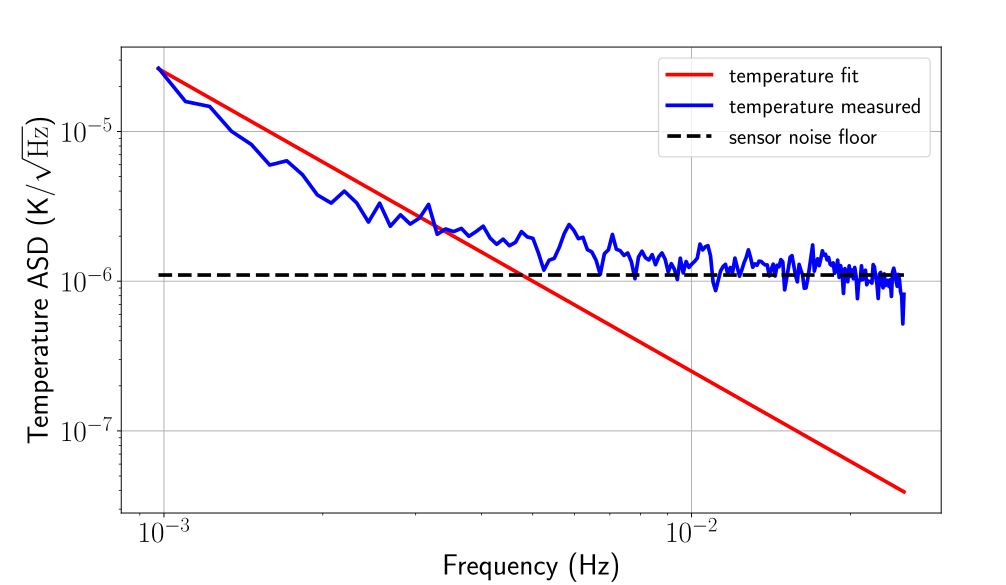}
\caption{The measured temperature ASD (solid blue) is fitted with a $1/f^2$ curve (solid red) for frequencies where the data are not dominated by sensor noise (dashed black).}
\label{fig:temp}
\end{figure}

Custom-made in-vacuum temperature sensors were used to estimate the temperature fluctuations at low frequencies. The temperature measurement is sensor noise-dominated above a few mHz. In general, temperature gradients can be assumed to decrease in magnitude with $\sim 1/f$ towards higher frequencies and are additionally low-pass filtered by the vacuum system, so that, above 1\,mHz, temperature fluctuations decrease with $\sim 1/f^2$ towards higher frequencies. Similar assumptions have been in made, for example, in \cite{Mow-Lowry2019}. Extrapolating the meaningful data by applying a fit curve with $1/f^2$ slope to the measured value at 1\,mHz leads to the temperature fluctuations being described by
\begin{eqnarray}
\label{eq:temp-fluctuations}
\Delta T\left(f\right) 
\approx 2.50 \times 10^{-11}\frac{\rm{K}}{\sqrt{\rm{Hz}}}\cdot \frac{1}{f^2}\,.
\end{eqnarray}
The measured temperature amplitude spectral density (ASD) and the applied fit are shown in \cref{fig:temp}.

\section{Alternative Suspension Designs}
\label{ap:susp-designs}
\begin{figure}[h]
\centering
\includegraphics[trim = 0cm 0cm 0cm 0cm, clip, width=13cm]{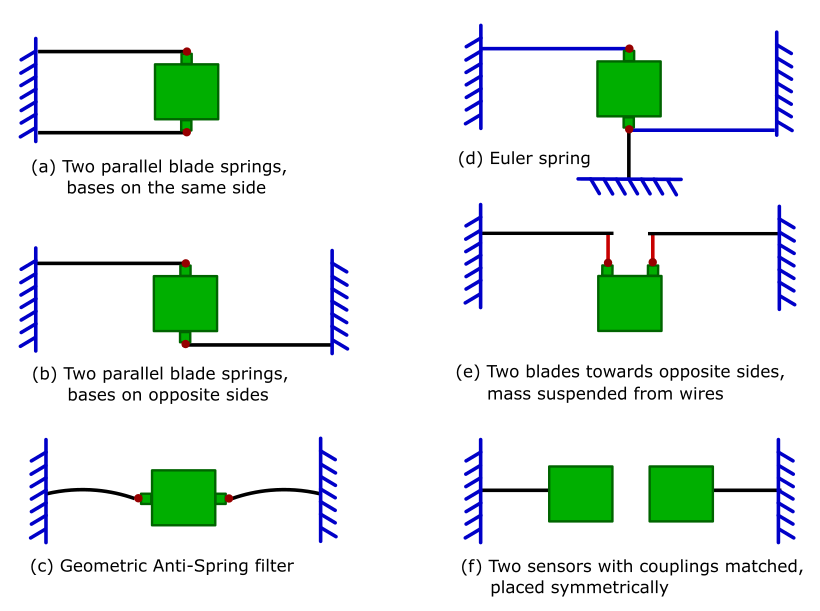}
\caption{Ideas for alternative suspension designs to reduce the strength of TTV coupling.}
\label{fig:susp-alternatives}
\end{figure}

In this section, the ideas in \cref{fig:susp-alternatives} for alternative suspension designs to overcome limitations of the single spring blade configuration are briefly discussed.\\
Configuration (a) suggests to use two parallel spring blades mounted on different heights of the test mass and having their bases on the same side. While this suspension is much stiffer in pitch and roll directions compared to a single blade spring and also decouples rotational motion from vertical motion of the test mass, the problem of TTV coupling is only solved for the roll direction whereas the mass still moves on a circle in pitch direction. This configuration also increases the fundamental vertical resonance which leads to a temperature drift reduction with $f_{0\rm{z}}^2$ but at the same time increases sensing noise with $f_{0\rm{z}}^{-2}$ as well as thermal noise with $f_{0\rm{z}}^{-1}$ towards lower frequencies.\\
The blade design for alternative (b) will probably be similar to the one for alternative (a). The only difference in the configuration is that the blade bases are on opposite sides which leads to the both blades pulling on the test mass for vertical deflection. Consequently, the centre of mass will move on a straight line and decouple the TTV coupling strength from drifts such that it is only dependent on the setup accuracy if the spring blades are equal.\\
In contrast to version (b), version (c) is quite a compact solution where the test mass moves on a straight line as well and, additionally, the alignment of the test mass is not fundamentally coupled to vertical motion. The disadvantage of using geometric anti-spring (GAS) filters \cite{Cella2005,Stochino2007} is that it complicates both design and assembly of the sensor.
 \\
For the use of Euler springs \cite{Winterflood2002} drawn schematically in part (d), the required mechanical stabilisation of the spring's tip would reintroduce the problem of the test mass moving on a circle which leads to TTV coupling. This problem could be avoided by using two flexures from opposite sides in accordance with configuration (b) but this reintroduces the fundamental coupling of angular alignment and vertical motion. Advantages of this configuration are that the fundamental resonance frequency can be pushed to low frequencies (sub-\SI{1}{\hertz} regime), fundamentally increasing the sensor sensitivity towards lower frequencies, and the extension of the blade in the vertical direction compactifying the setup.\\
Configuration (e) comprises two blades with their blade bases on opposite sides and additionally wires suspending the test mass. This is similar to suspensions of auxiliary optics in gravitational-wave detectors with low requirements and will have by far the lowest resonance frequencies for all degrees of freedom compared to the other configurations. Damping of horizontal and rotational degrees of freedom will probably be useful to prevent cross coupling. Also for this suspension design, the test mass would move on a straight line parallel to gravity such that TTV coupling will not depend on drifts but only on setup accuracy if the blades are equal. Furthermore, there is no fundamental cross coupling from other DoFs into vertical motion but only if the blades are twisted.\\
Finally, configuration (f) relies on subtracting the coherent tilt coupling of two sensors which are placed symmetrically with respect to the side and vertical axis of the optical table rather than changing the suspension design. A sufficient  subtraction of the unwanted signal requires that the tilt coupling strengths of both sensors are matched. This can be done by matching the static angular alignment of one sensor to the other so that they have the same dip frequency in their pitch transfer functions. The method accounts for the fundamental pitch coupling of the single blade spring suspension used for the results in this paper but does not help with possibly occurring coupling of roll motion into the sensor signal. The principle is equivalent to that of symmetric triaxial seismometers like the Streckeisen STS-2 where the orientation of the vertical readout would not change with temperature drifts, even if there was no force-feedback. It would also allow to keep working with other common designs like the leaf-spring seismometer \cite{Wielandt1982} which might otherwise be affected by drift-induced TTV coupling at some point.

\setcounter{section}{1}
\newpage
\bibliographystyle{iopart-num}
\bibliography{literature}

\end{document}